\documentclass[11pt]{article}

\usepackage[tbtags]{amsmath}
\usepackage{amsfonts,amssymb}

\begin{document}

\newcommand\ptl{\partial}
\newcommand\tl{\tilde}
\newcommand\mtr{\mathrm{matter}}
\newcommand\fm{\mathrm{fm}}
\newcommand\ms{\mathrm{ms}}
\newcommand\rf{\mathrm{f}}
\newcommand\cH{\mathcal{H}}
\newcommand\cL{\mathcal{L}}
\newcommand\cM{\mathcal{M}}


\title{ \vspace{-1.2in} Universally Coupled Massive Gravity\footnote{Published in \emph{Theoretical and Mathematical Physics} {\bf 151} no. 2, pp. 700-717 (2007). }}

\author{ \vspace{-.2in} J.~B.~Pitts\footnote{Department of Physics, University of Texas at Austin, Austin,
Texas, USA; and
Department of Mathematics, St.~Edward's University, Austin,
Texas; History and Philosophy of Science Grad\-uate Program,
University of Notre Dame, Notre Dame, Indiana, USA, e-mail:
jpitts@nd.edu.} and
W.~C.~Schieve\footnote{Department of Physics, University of Texas at Austin, Austin,
Texas, USA.} }

\date{ \vspace{-.5in} }


\maketitle

\begin{abstract}
We derive Einstein's equations from a linear theory in flat space-time using
free-field gauge invariance and universal coupling. The gravitational
potential can be either covariant or contravariant and of almost any density
weight. We adapt these results to yield universally coupled massive variants
of Einstein's equations, yielding two one-parameter families of distinct
theories with spin~$2$ and spin~$0$. The Freund-Maheshwari-Schonberg theory
is therefore not the unique universally coupled massive generalization of
Einstein's theory, although it is privileged in some respects. The theories
we derive are a subset of those found by Ogievetsky and Polubarinov by other
means. The question of positive energy, which continues to be discussed,
might be addressed numerically in spherical symmetry. We briefly comment on
the issue of causality with two observable metrics and the need for gauge
freedom and address some criticisms by Padmanabhan of field derivations of
Einstein-like equations along the way.
\end{abstract}


\section{Introduction}
\label{sec1}

Constructing a relativistic gravitational theory based on principles such as
an analogy to Maxwellian electromagnetism, the universal coupling of the
gravitational field to a combined gravity--matter energy--momentum complex,
and also the requirement that the gravitational field equations alone
(without the matter equations, again in analogy with electromagnetic charge
conservation) entail energy--momentum conservation was a major part of
Einstein's search for an adequate theory of gravity in 1913--1915
(see~\cite{1},~\cite{2}). Einstein subsequently downplayed these
investigations~\cite{2}, and the above ideas later came to be associated with
the non-Einsteinian field theory approach to gravitation. Such a derivation
of Einstein's or similar gravitational field equations can use a priori
preferred coordinates and a canonical energy--momentum tensor or a flat
background metric and variational metric energy--momentum tensor, the
difference being basically formal.

Several authors~\cite{3}--\cite{14} have discussed the utility of a flat
background metric $\eta_{\mu\nu}$ in general relativity or the possibility of
deriving that theory, approximately or exactly, from a flat space--time
theory.\footnote{We note that general relativity was also derived from
self-interaction on curved backgrounds~\cite{10},~\cite{15}. The possibility
(or otherwise) of massless multigraviton theories analogous to Yang--Mills
theories was also investigated~\cite{16}. Moreover, it is interesting to
consider (A)dS backgrounds, Lorentz-violating theories, and Chern--Simons
topologically massive theories. But we consider only the most traditional
form of the massive gravity problem here. We work in four space--time
dimensions, but the derivations should generalize straightforwardly to any
(integral) dimension not less than three.} A background metric allows
introducing a gravitational energy--momentum tensor~\cite{17}, not merely a
pseudotensor. As a result, gravitational energy and momentum are independent
of the coordinates but dependent on the gauge~\cite{18}. If we want to regard
the background metric seriously as a property of space--time and not just
treat it as a useful fiction, then the relation between the effective curved
metric's null cone 
and that of the flat background must be considered. That
issue was addressed with some success for the massless case of Einstein's
equations~\cite{13}.

Preparatory to considering massive theories of gravity, we generalize our
derivation of Einstein's equations using gauge invariance for the free field
and universal coupling~\cite{12} to permit almost any density weight and
either covariant or contravariant valency for the symmetric rank-two
gravitational potential. This generality to some degree parallels that in
Kraichnan's classic work~\cite{4}, but our derivation has several
improvements and can be easily adapted to massive theories. The choices of
index position and density weight make no difference (after field
redefinitions) in the massless theories, but they yield distinct massive
theories here. Along the way, we address Padmanabhan's recent objections to
field derivations of Einstein's equations~\cite{19}.

Several authors recently discussed the subject of massive gravity with spin-2
and spin-0 components~\cite{20},~\cite{21}. While it is permissible simply to
postulate the nonlinear features (if any) of a mass term, it seems preferable
to find well-motivated theoretical principles to constrain such choices. Some
time ago, Ogievetsky and Polubarinov (OP) derived a two-parameter family of
massive variants of Einstein's equations~\cite{7}, which contains both our
one-parameter families (quantization was considered briefly in~\cite{22}).
Their derivation relied on gauge invariance (at least for the massless part
of the Lagrangian density) but not universal coupling. Instead, they imposed
a spin-limitation principle to exclude some degrees of freedom, many with the
wrong sign and thus negative energy, from the full nonlinear interacting
theory. The Freund--Maheshwari--Schonberg (FMS) massive theory was originally
derived using universal coupling with a canonical, not metric,
energy--momentum tensor~\cite{23}. While the FMS theory is recovered among
our results, it is not the unique universally coupled massive variant of
Einstein's equations, contrary to previous claims~\cite{23},~\cite{24}. Our
derivation using the metric energy--momentum tensor seems shorter and cleaner
than the FMS derivation using the canonical energy--momentum tensor. In a
future work, we will consider universal coupling using a tetrad, not metric,
formalism, thus obtaining additional universally coupled theories, and will
use the metric formalism to show that all the OP theories are universally
coupled. All known universally coupled theories correspond to the OP family
of theories; hence, their derivation and the metric universal coupling
derivation currently lead to coinciding results. The FMS theory, subsequently
adopted by Logunov and collaborators~\cite{11}, also faces questions
regarding positive energy and causality, on which we briefly comment in what
follows.

\section{Effectively geometric theories from universal coupling and gauge
invariance}
\label{sec2}

We previously derived Einstein's theory and other effectively geometric
theories using gauge-invariant free-field theories and universal
coupling~\cite{12}, which was significantly based on the work of
Kraichnan~\cite{4} and Deser~\cite{9}. Whereas the gravitational potential
was previously taken to be a symmetric rank-two covariant tensor field, we
now generalize this derivation to the case of a density of almost any weight
and either covariant or contravariant valency. Given the generality of
Kraichnan's derivation, it is not surprising that these generalizations again
yield only Einstein's and other effectively geometric theories. Nothing
especially novel is obtained for massless theories, but the derivations below
are adapted to massive gravity for the first time.

\subsection{Free-field action for a covariant tensor density potential}
\label{sec2.1}
For the massless theories, an initial infinitesimal invariance (up to a
boundary term) of the free gravitational action is assumed. For the
subsequent derivation of massive theories, the gauge freedom is broken by a
natural mass term algebraic in the fields, but the derivative terms retain
the gauge invariance.

Let $S_{\rf}$ be the action for a free symmetric tensor density
$\tl{\gamma}_{\mu\nu}$ (of density weight $-l$, where $l\ne1/2$) in
a space--time with a flat metric tensor $\eta_{\mu\nu}$ in arbitrary
coordinates. The torsion-free metric-compatible covariant derivative
is denoted by $\ptl_{\mu}$; hence, $\ptl_{\alpha}\eta_{\mu\nu}=0$.
It is convenient to use not the flat metric itself but its related
densitized metric
$\tl{\eta}_{\mu\nu}=\eta_{\mu\nu}(\sqrt{-\eta}\,)^{-l}$ of weight
$-l$. We note that in the forbidden case $l=1/2$,
$\tl{\eta}_{\mu\nu}$ is noninvertible:
$\eta_{\mu\nu}(\sqrt{-\eta}\,)^{-1/2}$ determines only the null
cone, not a full metric tensor.\footnote{This exceptional case was
of interest in deriving slightly bimetric theories~\cite{12}, where
$\sqrt{-\eta}$ can appear in the field equations of the interacting
theories.} The field $\tl{\gamma}_{\mu\nu}$ turns out to be the
gravitational potential. Although it has become customary in work on
loop quantum gravity to denote the density weight of fields by
placing the corresponding number of tildes above or below a symbol
to express its positive or negative density weight, that custom is
impossible here. The density weight $l$ ($l\ne1/2$) can be large,
nonintegral, or even irrational, and we therefore merely write a
tilde over most densities. All indices are respectively raised and
lowered with $\eta^{\mu\nu}$ and $\eta_{\mu\nu}$ with two
exceptions. For the densitized flat metric $\tl{\eta}_{\mu\nu}$, the
oppositely densitized inverse flat metric is $\tl{\eta}^{\mu\nu}$.
Similarly, the inverse of the densitized curved metric
$\tl{g}_{\mu\nu}$ defined below is $\tl{g}^{\mu\nu}$. Because we use
tensor densities extensively, we recall the forms of their covariant
and Lie derivatives. A $(1,1)$-density $\tl{\phi}^{\alpha}_{\beta}$
of weight $w$ is a representative example. The Lie derivative is
given by~\cite{25}
\begin{equation}
\pounds_{\xi}\tl{\phi}^{\alpha}_{\beta}=
\xi^{\mu}\tl{\phi}^{\alpha}_{\beta},_{\mu}-
\tl{\phi}^{\mu}_{\beta}\xi^{\alpha},_{\mu}+
\tl{\phi}^{\alpha}_{\mu}\xi^{\mu},_{\beta}+
w\tl{\phi}^{\alpha}_{\beta}\xi^{\mu},_{\mu},
\label{1}
\end{equation}
and the $\eta$-covariant derivative is given by
\begin{equation}
\ptl_{\mu}\tl{\phi}^{\alpha}_{\beta}=\tl{\phi}^{\alpha}_{\beta},_{\mu}+
\tl{\phi}^{\sigma}_{\beta}\Gamma_{\sigma\mu}^{\alpha}-
\tl{\phi}^{\alpha}_{\sigma}\Gamma_{\beta\mu}^{\sigma}-
w\tl{\phi}^{\alpha}_{\beta}\Gamma_{\sigma\mu}^{\sigma},
\label{2}
\end{equation}
where $\Gamma_{\beta\mu}^{\sigma}$ are the Christoffel symbols for
$\eta_{\mu\nu}$. After the curved metric $g_{\mu\nu}$ is defined below, the
analogous $g$-covariant derivative $\nabla$ with the Christoffel symbols
$\{_{\sigma\mu}^{\alpha}\}$ follows.

The desire to avoid ghosts motivates gauge invariance for linear
theories~\cite{26}. We require that the free field action $S_{\rf}$ change
only by a boundary term under the infinitesimal gauge transformation
\begin{equation}
\tl{\gamma}_{\mu\nu}\to
\tl{\gamma}_{\mu\nu}+\delta\tl{\gamma}_{\mu\nu},\qquad
\delta\tl{\gamma}_{\mu\nu}=\ptl_{\mu}\tl{\xi}_{\nu}+
\ptl_{\nu}\tl{\xi}_{\mu}+c\eta_{\mu\nu}\ptl^{\alpha}\tl{\xi}_{\alpha},
\label{3}
\end{equation}
where $c\ne-1/2$ and $\tl{\xi}_{\nu}$ is an arbitrary covector density field
of weight $-l$.\footnote{The case $c=-1/2$ merely gives a scalar theory in
the somewhat comparable work of OP~\cite{7}.} We can expect the appearance of
a connection between $l$ and $c$. For any $S_{\rf}$ invariant in this sense
under~\eqref{3}, a certain linear combination of the free field equations is
identically divergenceless, as we now show. The action changes by
\begin{equation}
\delta S_{\rf}=\int d^4x\left[\frac{\delta S_{\rf}}
{\delta\tl{\gamma}_{\mu\nu}}(\ptl_{\nu}\tl{\xi}_{\mu}+
\ptl_{\mu}\tl{\xi}_{\nu}+c\eta_{\mu\nu}\ptl^{\alpha}\tl{\xi}_{\alpha})+
e^{\mu},_{\mu}\right]=\int d^4x\, f^{\mu},_{\mu}.
\label{4}
\end{equation}
The explicit forms of the boundary terms given by $e^{\mu},_{\mu}$ and
$f^{\mu},_{\mu}$ are not needed for our purposes. Integrating by parts,
letting $\tl{\xi}_{\mu}$ have compact support to annihilate the boundary
terms, and using the arbitrariness of $\tl{\xi}_{\mu}$, we obtain the
identity
\begin{equation}
\ptl_{\mu}\left(\frac{\delta S_{\rf}}{\delta\tl{\gamma}_{\mu\nu}}+
\frac c2\eta^{\mu\nu}\eta_{\sigma\alpha}
\frac{\delta S_{\rf}}{\delta\tl{\gamma}_{\sigma\alpha}}\right)=0.
\label{5}
\end{equation}
This is the generalized Bianchi identity for the free theory, which in the
most common case is a linearized version of the original geometric Bianchi
identity. A natural choice for $S_{\rf}$ is the linearized GR Lagrangian
density, but no such detailed assumptions on the form of $S_{\rf}$ are used
here. Because noninteracting sourceless fields are unobservable~\cite{27},
the theory is interesting only after an interaction is introduced.

As Kretschmann pointed out long ago in response to Einstein, any theory can
be given a generally covariant formulation in the sense that the equations
hold in any coordinate system, Cartesian or otherwise~\cite{28}--\cite{30}.
The resulting formulation might be called weakly or trivially general
covariant. Often achieving weak general covariance involves using absolute
objects~\cite{29},~\cite{29a}, such as a flat metric tensor, as it does here.
The distinctively novel aspect of Einstein's theory of gravity is supposedly
its lack of absolute objects or ``prior
geometry"~\cite{29},~\cite{31}.\footnote{Actually, the matter is more
complicated: R.~Geroch and Giulini recently noted in effect that $g$, the
determinant of the metric tensor, is an absolute object because any point has
a neighborhood with a coordinate system such that the component of $g$ has
the value $-1$~\cite{32},~\cite{32a}.} This property might be called
strong or nontrivial general covariance. This distinction between two senses
of general covariance was discussed previously~\cite{33}. The derivation
presented here and in some other works starts with a weakly generally
covariant theory and concludes with a strongly generally covariant one
without absolute objects.\footnote{The fact that $g$ counts as absolute in
general relativity in the Anderson--Friedman absolute-object program suggests
that general relativity is not strongly covariant after all. But the point
remains that two very different notions of general covariance are in play. In
addition to the field equations, the topology, boundary conditions, and
causality should be examined in seeking absolute objects~\cite{13}. We
emphasize that the massless cases are considered here; obviously, the massive
theories contain the flat metric and are therefore not strongly generally
covariant.} Therefore, the quite misleading claim~\cite{19} that flat
space--time derivations of Einstein's equations only result in general
covariance because they feed it in at the beginning should not be accepted.
The two senses of general covariance are very different; Padmanabhan's
criticism commits the fallacy of equivocation.

\subsection{Metric energy--momentum tensor density}
\label{sec2.2}
If the energy--momentum tensor is to be the source of the gravitational
potential $\tl{\gamma}_{\mu\nu}$, then consistency requires that the {\sl
total} energy--momentum tensor be used including the gravitational
energy--momentum, not merely the nongravitational (``matter")
energy--momentum, because only the total energy--momentum tensor is
divergenceless in the sense of $\ptl_{\nu}$~\cite{9} or, equivalently, in the
sense of a Cartesian coordinate divergence. To obtain a global conservation
law, a vanishing {\sl coordinate} divergence for the four-current is needed.

An expression for the total energy--momentum tensor density can be derived
from $S$ using the metric recipe~\cite{4},~\cite{17},~\cite{29},~\cite{34} as
follows. The action depends on the flat metric density $\tl{\eta}_{\mu\nu}$,
gravitational potential $\tl{\gamma}_{\mu\nu}$, and matter fields $u$. Here,
$u$ represents an arbitrary collection of dynamical tensor fields of
arbitrary rank, index position, and density weight. Using the OP spinor
formalism that uses the ``square root of the metric" rather than a tetrad or
other additional structure~\cite{35}, we can likely also include spinor
fields.

Under an arbitrary infinitesimal coordinate transformation described by a
vector field $\xi^{\mu}$, the action changes by the amount
\begin{equation}
\delta S=\int d^4x\left(\frac{\delta S}{\delta\tl{\gamma}_{\mu\nu}}
\pounds_{\xi}\tl{\gamma}_{\mu\nu}+\frac{\delta S}{\delta u}\pounds_{\xi}u+
\frac{\delta S}{\delta\tl{\eta}_{\mu\nu}}\pounds_{\xi}
\tl{\eta}_{\mu\nu}+g^{\mu},_{\mu}\right),
\label{6}
\end{equation}
with boundary terms from $g^{\mu},_{\mu}$ vanishing because $\xi^{\mu}$ is
compactly supported. But $S$ is a scalar, and hence $\delta S=0$. Letting the
matter and gravitational field equations hold, integrating by parts,
discarding vanishing boundary terms, and using the arbitrariness of the
vector field $\xi^{\mu}$ gives
\begin{equation}
\ptl_{\nu}\left(\frac{\delta S}{\delta\tl{\eta}_{\mu\nu}}-
\frac l2\tl{\eta}_{\alpha\beta}\tl{\eta}^{\mu\nu}
\frac{\delta S}{\delta\tl{\eta}_{\alpha\beta}}\right)=0.
\label{7}
\end{equation}
This quantity is an energy--momentum tensor density for matter and
gravitational fields. The flatness of $\eta_{\mu\nu}$ is relaxed in taking
the functional derivative $\delta S/\delta\tl{\eta}_{\mu\nu}$ and is restored
later. This move is sometimes criticized~\cite{19}, but it is unobjectionable
even in flat space--time because it is merely a formal trick useful for
defining the energy--momentum tensor, not an illicit use of curved
space--time. Using the connection between the Rosenfeld energy--momentum
tensor and the symmetrized Belinfante canonical energy--momentum
tensor~\cite{34},~\cite{36}, we could regard the metric recipe as a
mathematical shortcut to the conceptually unimpeachable but mathematically
inconvenient Belinfante tensor modified with terms proportional to the
equations of motion. The metric energy--momentum tensor is not unique at this
stage, because terms proportional to the equations of motion and their
derivatives, as well as superpotentials, can be added. Another option would
be to use Lagrange multipliers and let the background metric be flat only
on-shell~\cite{37}. As is shown below, using the superpotential freedom
judiciously is important in deriving the field equations.

\subsection{Full universally coupled action}
\label{sec2.3}
We find an action $S$ satisfying the plausible physical postulate that
invertible linear combinations of the Euler--Lagrange equations are just
invertible linear combinations of the free field equations for $S_{\rf}$
augmented by the total energy--momentum tensor. A simple way to impose this
requirement reduces to the condition
\begin{equation}
\frac{\delta S}{\delta\tl{\gamma}_{\mu\nu}}=
\frac{\delta S_{\rf}}{\delta\tl{\gamma}_{\mu\nu}}-
\lambda\frac{\delta S}{\delta\tl{\eta}_{\mu\nu}},
\label{8}
\end{equation}
where $\lambda=-\sqrt{32\pi G}$. (Because the linear combinations of the free
Euler--Lagrange equations and of the full Lagrange equations with interaction
are invertible in this case, the linear combination can now be applied to the
energy--momentum tensor.) It seems prudent to set $c=-l$ to make the
generalized Bianchi identity and the energy--momentum tensor density take
similar forms.

The basic variables here are the gravitational potential
$\tl{\gamma}_{\mu\nu}$ and the flat metric density $\tl{\eta}_{\mu\nu}$. But
we can freely change the variables in $S$ from $\tl{\gamma}_{\mu\nu}$ and
$\tl{\eta}_{\mu\nu}$ to the bimetric variables $\tl{g}_{\mu\nu}$ and
$\tl{\eta}_{\mu\nu}$~\cite{4}, where
\begin{equation}
\tl{g}_{\mu\nu}=\tl{\eta}_{\mu\nu}-\lambda\tl{\gamma}_{\mu\nu}.
\label{9}
\end{equation}
(We can then define the metric $g_{\mu\nu}$ from $\tl{g}_{\mu\nu}$ using
matrix algebra and then define the $g$-covariant derivative $\nabla$ as
usual, but we have little need to use $\nabla$ explicitly.)

Equating coefficients of the variations gives
\begin{equation}
\frac{\delta S}{\delta\tl{\eta}_{\mu\nu}}\bigg|\tl{\gamma}=
\frac{\delta S}{\delta\tl{\eta}_{\mu\nu}}\bigg|\tl{g}+
\frac{\delta S}{\delta\tl{g}_{\mu\nu}}
\label{10}
\end{equation}
for $\delta\tl{\eta}_{\mu\nu}$ and
\begin{equation}
\frac{\delta S}{\delta\tl{\gamma}_{\mu\nu}}=
-\lambda\frac{\delta S}{\delta\tl{g}_{\mu\nu}}
\label{11}
\end{equation}
for $\delta\tl{\gamma}_{\mu\nu}$. Combining these two results gives
\begin{equation}
\lambda\frac{\delta S}{\delta\tl{\eta}_{\mu\nu}}\bigg|\tl{\gamma}=
\lambda\frac{\delta S}{\delta\tl{\eta}_{\mu\nu}}\bigg|\tl{g}-
\frac{\delta S}{\delta\tl{\gamma}_{\mu\nu}}.
\label{12}
\end{equation}
Equation~\eqref{12} splits the energy--momentum tensor into two parts: one
that vanishes when gravity is on-shell and one that does not. Using this
result in universal-coupling postulate~\eqref{8} gives
\begin{equation}
\lambda\frac{\delta S}{\delta\tl{\eta}_{\mu\nu}}\bigg|\tl{g}=
\frac{\delta S_{\rf}}{\delta\tl{\gamma}_{\mu\nu}}.
\label{13}
\end{equation}

Taking the divergence, recalling free-theory Bianchi identity~\eqref{5}, and
using $c=-l$, we derive
\begin{equation}
\ptl_{\mu}\left(\frac{\delta S}{\delta\tl{\eta}_{\mu\nu}}\bigg|\tl{g}-
\frac l2\tl{\eta}^{\mu\nu}\tl{\eta}_{\alpha\beta}
\frac{\delta S}{\delta\tl{\eta}_{\alpha\beta}}\bigg|\tl{g}\right)=0.
\label{14}
\end{equation}
The quantity in parentheses is exactly $(\sqrt{-\eta}\,)^l\,
\delta S/\delta\eta_{\mu\nu}|\tl{g}$. Hence, the part of the energy--momentum
tensor not proportional to the gravitational field equations has identically
vanishing divergence (on either index), i.e., is a (symmetric)
``curl"~\cite{29}. The splitting of the energy--momentum tensor ensures that
in the massless case, the gravitational field equations {\sl alone}, without
separately postulating the matter equations, entail conservation of
energy--momentum for the resulting effectively geometric field equations.

Because the quantity $\delta S/\delta\eta_{\mu\nu}|\tl{g}$ is symmetric and
has identically vanishing divergence on either index, it necessarily has the
form~\cite{38}
\begin{equation}
\frac{\delta S}{\delta\eta_{\mu\nu}}\bigg|\tl{g}=
\frac12\ptl_{\rho}\ptl_{\sigma}\bigl(\cM^{[\mu\rho][\sigma\nu]}+
\cM^{[\nu\rho][\sigma\mu]}\bigr)+B\sqrt{-\eta}\eta^{\mu\nu},
\label{15}
\end{equation}
where $\cM^{\mu\rho\sigma\nu}$ is a tensor density of unit weight and $B$ is
a constant. This result follows from the converse of the Poincar\'e lemma in
Minkowski space--time. We cannot choose $\cM ^{\mu\rho\sigma\nu}$
arbitrarily but must choose it such that the term $\delta S_{\rf}/
\delta\tl{\gamma}_{\mu\nu}$ is accommodated. The freedom to add an arbitrary
curl must therefore be used in a quite definite way. As Huggins, a student of
Feynman, showed in his dissertation~\cite{39} and Padmanabhan recently
emphasized~\cite{19}, coupling a spin-2 field to the energy--momentum tensor
does not lead to a unique theory, because of terms of this curl form. Rather,
as Huggins said (p.~39 in~\cite{39}): ``an additional restriction is
necessary. For Feynman this restriction was that the equations of motion be
obtained from an action principle; Einstein required that the gravitational
field have a geometric interpretation. Feynman showed these two restrictions
to be equivalent."

Gathering all dependence on $\eta_{\mu\nu}$ (with $\tl{g}_{\mu\nu}$
independent) into one term yields $S=S_1[\tl{g}_{\mu\nu},u]+
S_2[\tl{g}_{\mu\nu},\eta_{\mu\nu},u]$. It is easy to verify that if
\begin{equation}
S_2=\frac12\int d^4x\,R_{\mu\nu\rho\sigma}(\eta)
\cM^{\mu\nu\rho\sigma}(\eta_{\mu\nu},\tl{g}_{\mu\nu},u)+
\int d^4x\,\alpha^{\mu},_{\mu}+2B\int d^4x\,\sqrt{-\eta},
\label{16}
\end{equation}
then $\delta S_2/\delta\eta_{\mu\nu}|\tl{g}$ has exactly the desired form,
while $S_2$ does not affect the Euler--Lagrange equations, because
$\delta S_2/\delta\tl{g}_{\mu\nu}=0$ and $\delta S_2/\delta u=0$
identically~\cite{4}.\footnote{If it seems that using this term $(1/2)\int
d^4x\,R_{\mu\nu\rho\sigma}(\eta)\cM^{\mu\nu\rho\sigma}(\eta_{\mu\nu},
\tl g_{\mu\nu},u)$ is too clever to be invented without knowing Einstein's
theory in advance and thus cheating~\cite{19}, then a symmetric curl term
$\ptl_{\rho}\ptl_{\sigma}(\cM^{[\mu\rho][\sigma\nu]}+
\cM^{[\nu\rho][\sigma\mu]})/2+B\sqrt{-\eta}\eta^{\mu\nu}$ should be simply
added to the metric energy--momentum tensor by hand using the usual
underdetermination of the energy--momentum tensor.} The coefficient $B$ of
the four-volume term is naturally chosen to cancel any zeroth-order term
(such as from a cosmological constant) in the action such that the action
vanishes when there is no gravitational field. The four-divergence
$\alpha^{\mu},_{\mu}$ resolves worries~\cite{19} about obtaining terms that
are not analytic in the coupling constant $\lambda$. It is unclear whether
the Hilbert action is best in any event, given its badly behaved conservation
laws~\cite{40}.

Hence, the universally coupled action for the covariant tensor density case
is
\begin{equation}
S=S_1 [\tl{g}_{\mu\nu},u]+\frac12\int d^4x\,R_{\mu\nu\rho\sigma}(\eta)
\cM^{\mu\nu\rho\sigma}+2B\int d^4x\,\sqrt{-\eta}+
\int d^4x\,\alpha^{\mu},_{\mu}.
\label{17}
\end{equation}
The boundary term is at our disposal; if $\alpha^{\mu}$ is a unit-weight
vector density, then $S$ is a coordinate scalar. Using the effective curved
metric density $\tl{g}_{\mu\nu}$, we can define an effective curved metric by
$\tl{g}_{\mu\nu}=g_{\mu\nu}(\sqrt{-g}\,)^{-l}$ and an inverse curved metric
density $\tl{g}^{\mu\nu}=g^{\mu\nu}(\sqrt{-g}\,)^l$.

For $S_1$, we choose the Hilbert action for general relativity plus
minimally coupled matter and a cosmological constant:
\begin{equation}
S_1=\frac1{16\pi G}\int d^4x\,\sqrt{-g}\,R(g)-
\frac{\Lambda}{8\pi G}\int d^4x\,\sqrt{-g}+S_{\mtr}[g_{\mu\nu},u].
\end{equation}
It is well known that the Hilbert action is the simplest (scalar) action that
can be constructed using only the metric tensor. If the gravitational field
vanishes everywhere, then the gravitational action should also vanish. In the
massless case, the result is that $B=\Lambda/(16\pi G)$. For the
generalization to the massive case considered below, the gauge-breaking part
of the mass term introduces another zeroth-order contribution that also needs
to be canceled. It is also possible to couple the matter to the Riemann
tensor for $g_{\mu\nu}$ or to allow higher powers of the Riemann tensor in
the gravitational action, if desired. In the massless case, we could set
$\Lambda=0$~\cite{23}.

\section{Massive universally coupled gravity for a covariant tensor density
potential}
\label{sec3}

Our goal is to generalize the above derivation to obtain one or more massive
finite-range variants of Einstein's equations. Such field equations would be
related to Einstein's much as Proca's massive electromagnetic field equations
are related to Maxwell's. But a spin-2--spin-0 massive theory would have
ghosts at the level of the free linear theory. Good linear behavior is
generally required as a guide to good nonlinear behavior. But good linear
behavior seems neither necessary nor sufficient for good nonlinear behavior.
In the present case, it seems quite possible that the nonlinear form of the
Hamiltonian constraint cures the bad behavior of the linear theory. We
briefly discuss this matter below.

It can be expected that the mass term for a free field is quadratic in the
potential and lacks derivatives. The free-field action $S_{\rf}$ is now
assumed to have two parts: a (mostly kinetic) part $S_{\rf0}$ that is
invariant under the previous gauge transformations as in the massless case
above and an algebraic mass term $S_{\fm}$ that is quadratic and breaks the
gauge symmetry. We seek a full universally coupled theory with an action $S$
that has two corresponding parts: $S=S_0+S_{\ms}$. They are the familiar part
$S_0$ (yielding the Einstein tensor, the matter action, a cosmological
constant, and a zeroth-order four-volume term) and the new gauge-breaking
part $S_{\ms}$, which also has another zeroth-order four-volume term. As it
turns out, the mass term is constructed from both the algebraic part of $S_0$
(the cosmological constant and four-volume term) and the purely algebraic
term $S_{\ms}$.

Requiring $S_{\rf0}$ to change only by boundary terms under the variation
$\tl{\gamma}_{\mu\nu}\to\tl{\gamma}_{\mu\nu}+\ptl_{\mu}\tl{\xi}_{\nu}+
\ptl_{\nu}\tl{\xi}_{\mu}+c\eta_{\mu\nu}\ptl^{\alpha}\tl{\xi}_{\alpha}$ for
$c\ne-1/2$ implies the identity
\begin{equation}
\ptl_{\mu}\left(\frac{\delta S_{\rf0}}{\delta\tl{\gamma}_{\mu\nu}}-
\frac l2\eta^{\mu\nu}\eta_{\sigma\alpha}
\frac{\delta S_{\rf0}}{\delta\tl{\gamma}_{\sigma\alpha}}\right)=0.
\label{19}
\end{equation}
We again postulate the universal coupling in the form
\begin{equation}
\frac{\delta S}{\delta\tl{\gamma}_{\mu\nu}}=
\frac{\delta S_{\rf}}{\delta\tl{\gamma}_{\mu\nu}}-
\lambda\frac{\delta S}{\delta\tl{\eta}_{\mu\nu}}.
\label{20}
\end{equation}
Changing to the bimetric variables $\tl{g}_{\mu\nu}$ and
$\tl{\eta}_{\mu\nu}$, as before, implies that
\begin{equation}
\frac{\delta S_{\rf}}{\delta\tl{\gamma}_{\mu\nu}}=
\lambda\frac{\delta S}{\delta\tl{\eta}_{\mu\nu}}\bigg|\tl{g}.
\label{21}
\end{equation}

We now introduce the quantities $S_{\fm}$ and $S_{\ms}$, $S_{\rf}=S_{\rf0}+
S_{\fm}$ and $S=S_0+S_{\ms}$, in order to treat the pieces that existed in
the massless case separately from the new pieces in the massive case. We thus
obtain
\begin{equation}
\frac{\delta S_{\rf0}}{\delta\tl{\gamma}_{\mu\nu}}+
\frac{\delta S_{\fm}}{\delta\tl{\gamma}_{\mu\nu}}=
\lambda\frac{\delta S_0}{\delta\tl{\eta}_{\mu\nu}}\bigg|\tl{g}+
\lambda\frac{\delta S_{\ms}}{\delta\tl{\eta}_{\mu\nu}}\bigg|\tl{g}.
\label{22}
\end{equation}
Assuming that the new terms $S_{\fm}$ and $S_{\ms}$ correspond, we separate
this equation into the familiar part $\delta S_{\rf0}/
\delta\tl{\gamma}_{\mu\nu}=\lambda\,(\delta S_0/\delta\tl{\eta}_{\mu\nu})
|\tl{g}$ and the new part $\delta S_{\fm}/\delta\tl{\gamma}_{\mu\nu}=
\lambda\,(\delta S_{\ms}/\delta\tl{\eta}_{\mu\nu})|\tl{g}$. Using
invariance~\eqref{19}, we derive the form of $S_0$ as
\begin{equation}
S_0=S_1[\tl{g}_{\mu\nu},u]+S_2,
\label{23}
\end{equation}
as in the massless case. We again choose the simplest case and obtain the
Hilbert action with a cosmological constant, with matter coupled only to the
curved metric.

The new part in the massive case is
$$
\frac{\delta S_{\fm}}{\delta\tl{\gamma}_{\mu\nu}}=
\lambda\frac{\delta S_{\ms}}{\delta\tl{\eta}_{\mu\nu}}\bigg|\tl{g}.
$$
Assuming that the free-field mass term is quadratic in the gravitational
potential, we find that its variational derivative is
$$
\frac{\delta S_{\fm}}{\delta\tl{\gamma}_{\mu\nu}}=
a\sqrt{-\eta}\,\tl{\gamma}_{\alpha\beta}
(\tl{\eta}^{\alpha\mu}\tl{\eta}^{\beta\nu}+
b\tl{\eta}^{\alpha\beta}\tl{\eta}^{\mu\nu}).
$$
Changing to the bimetric variables gives
\begin{equation}
\frac{a\sqrt{-\eta}}{\lambda}(-\tl{g}_{\alpha\beta}+
\tl{\eta}_{\alpha\beta})(\tl{\eta}^{\alpha\mu}\tl{\eta}^{\beta\nu}+
b\tl{\eta}^{\alpha\beta}\tl{\eta}^{\mu\nu})=
\lambda\frac{\delta S_{\ms}}{\delta\tl{\eta}_{\mu\nu}}\bigg|\tl{g}.
\label{24}
\end{equation}
We take the expression for $S_{\ms}$ in the natural form
\begin{equation}
S_{\ms}=\int d^4x\,(p\tl{g}_{\alpha\beta}
\tl{\eta}^{\alpha\beta}+q)\sqrt{-\eta},
\label{24a}
\end{equation}
where $p$ and $q$ are real numbers to be determined below. We note that the
term $\sqrt{-g}$ itself, which gives a cosmological constant, plays no role
here and is already included in $S_0$. Using the relation
$\sqrt{-\eta}=(\sqrt{-\tl{\eta}}\,)^{1/(1-2l)}$, we calculate
$(\delta S_{\ms}/\delta\tl{\eta}_{\mu\nu})|\tl{g}$. Equating
$\lambda\,(\delta S_{\ms}/\delta\tl{\eta}_{\mu\nu})|\tl{g}$ to
$\delta S_{\fm}/\delta\tl{\gamma}_{\mu\nu}$, i.e., equating corresponding
coefficients, determines several of the constants. Equating the coefficients
of the $\sqrt{-\eta}\,\tl{\eta}_{\mu\nu}$ terms gives
$q=(2-4l)a(1+4b)/\lambda^2$, equating the coefficients of the
$\sqrt{-\eta}\,\tl{\eta}^{\mu\nu}\tl{\eta}^{\alpha\beta}\tl{g}_{\alpha\beta}$
terms gives $p=-ab(2-4l)/\lambda^2$, and equating the coefficients of the
$\sqrt{-\eta}\,\tl{\eta}^{\alpha\mu}\tl{\eta}^{\beta\nu}\tl{g}_{\alpha\beta}$
terms gives $p=a/\lambda^2$. Using the last two results together gives
$b=1/(4l-2)$. Using all three results together gives $q=-2a(2l+1)/\lambda^2$.
Combining the algebraic piece of $S_0$ with $S_{\ms}$ gives
\begin{equation}
S_{\mathrm{alg}}=-\frac{\Lambda}{8\pi G}\int d^4x\,\sqrt{-g}+
2B\int d^4x\,\sqrt{-\eta}+\frac a{\lambda^2}\int d^4x\,(\tl{g}_{\alpha\beta}
\tl{\eta}^{\alpha\beta}-4l-2)\sqrt{-\eta}.
\label{25}
\end{equation}

When the gravitational potential vanishes, $S_{\mathrm{alg}}$ and hence the
zeroth-order term should also vanish. Imposing this condition gives
$$
B=\frac{\Lambda}{16\pi G}-\frac{a(1-2l)}{\lambda^2}.
$$
Because our goal is to find a massive generalization of Einstein's theory,
not a theory with an effective cosmological constant, we require that the
first-order term in $\tl{\gamma}_{\mu\nu}$ also vanish. Because $\lambda^2=
32\pi G$, it follows that $\Lambda=a(1-2l)/2$. Hence, the sign of the
{\sl formal} cosmological-constant term depends on the density weight of the
initially chosen potential. We also expect the quadratic part of the
algebraic component $S_{\mathrm{alg}}$ of the action to agree with the
free-field mass term $S_{\fm}$. After a binomial expansion and some algebra,
we see that this is the case.\footnote{In~\cite{24}, there is a mistake in
the binomial expansion for $\sqrt{-g}$ between Eqs.~\thetag{43}
and~\thetag{44}.} The weak-field expansion of the full massive nonlinear
action $S$ allows relating the coefficient $a$ to the mass $m$ of the spin-2
gravitons: $a=-m^2$. For nontachyonic theories, we impose the condition
$a<0$.

Combining all these results gives the total massive action $S$, which depends
on the spin-2 graviton mass $m$ and the parameter $l$ controlling the
relative mass of the spin-0 ghost:
\begin{align}
S={}&\frac1{16\pi G}\int d^4x\,\sqrt{-g}\,R(g)+S_{\mtr}[\tl{g}_{\mu\nu},u]+{}
\nonumber
\\[2mm]
&+\frac12\int d^4x\,R_{\mu\nu\rho\sigma}(\eta)
\cM^{\mu\nu\rho\sigma}[\tl{\eta}_{\mu\nu},\tl{g}_{\mu\nu},u]+
\int d^4x\,\alpha^{\mu},_{\mu}-{}
\nonumber
\\[2mm]
&-\frac{m^2}{16\pi G}\int d^4x\,\biggl[\sqrt{-g}(2l-1)-\sqrt{-\eta}(2l+1)+
\frac12\sqrt{-\eta}\,\tl{g}_{\mu\nu}\tl{\eta}^{\mu\nu}\biggr],
\label{26}
\end{align}
where $l\ne1/2$. These theories are all universally coupled, contrary to the
claim that only the FMS theory has this property~\cite{23},~\cite{24}.

The Euler--Lagrange equations are easily found if the metric $g_{\mu\nu}$ is
used as the dynamical variable. The result is
\begin{equation}
\frac{\delta S}{\delta g_{\mu\nu}}=-\frac{\sqrt{-g}}{16\pi G}G^{\mu\nu}-
\frac{m^2}{16\pi G}\biggl[\frac{2l-1}2\sqrt{-g}\,g^{\mu\nu}+
\frac{(\sqrt{-\eta}\,)^{l+1}}{4(\sqrt{-g}\,)^l}
(2\eta^{\mu\nu}-l\eta^{\alpha\beta}g_{\alpha\beta}g^{\mu\nu})\biggr]+
\frac{\delta S_{\mtr}}{\delta g_{\mu\nu}}.
\label{27}
\end{equation}
Following Boulware and Deser~\cite{24}, we can linearize these theories and
check whether the spin-0 component is tachyonic. Nontachyonicity entails
$-1/2\le l<1/2$. For $l=0$, the spin-0 ghost has the same mass as the spin-2
degrees of freedom, and this theory (with density weight zero for the
potential) is hence the cleanest in the set. (The connection between density
``weight" and graviton ``mass" is an amusing linguistic accident.)
Investigating various masses for the spin-0 degree of freedom might have some
empirical consequences. This is especially the case in large-scale and
homogeneous situations, for example, in cosmology~\cite{21}. The ratio of the
spin-0 mass $m_0$ to the spin-2 mass $m_2$ is given by
\begin{equation}
\frac{m_0^2}{m_2^2}=\frac{-4l^2+1}{2l^2+1}.
\label{28}
\end{equation}
Hence, the mass is an even function of $l$. For $l=-1/2$, the spin-0 degree
of freedom is massless, and $\sqrt{-\eta}$ is absent from the mass term. As
$l\to1/2$, the scalar again becomes light, and the coefficient of $\sqrt{-g}$
tends to zero, although the value $l=1/2$ is forbidden. Between these
massless endpoints, the spin-0 degree of freedom becomes heavier. At the
midpoint $l=0$, the scalar has the same mass as the spin-2 field, giving a
simple form of the wave equation for the linearized massive Einstein
equations. Thus, the spin-0 ghost is never heavier than the spin-2 degrees of
freedom, and the $l{\ne}0$ theories hence have weaker gravitational
attraction at large distances and in homogeneous situations.

\section{Derivation of a contravariant tensor density potential: Massless
case}
\label{sec4}

We briefly present the contravariant analogue of the above derivation of
Einstein's equations. The gravitational potential is now a contravariant
symmetric tensor density field $\tl{\gamma}^{\mu\nu}$ of density weight $l$,
where $l\ne1/2$. In addition to the obvious moves of some indices, we
introduce some sign changes in the contravariant case. The hope for a simple
rule relating index moves and sign changes is disappointed because the
symmetry between the $(0,2)$ theory of weight $-l$ and the $(2,0)$ theory of
weight $l$ is broken: the Lagrangian density is a scalar density of weight~1,
not weight~0. The consequences of this imperfect symmetry are more apparent
in the massive case below than in the current massless case. It is convenient
to use not the inverse flat metric itself but the densitized related metric
$\tl{\eta}^{\mu\nu}=\eta^{\mu\nu}(\sqrt{-\eta}\,)^l$ of weight $l$, where
$l\ne1/2$.

In the massless theories, we assume an initial infinitesimal invariance (up
to a boundary term) of the free gravitational action $S_{\rf}$ under the
infinitesimal gauge transformation $\tl{\gamma}^{\mu\nu}\to
\tl{\gamma}_{\mu\nu}+\delta\tl{\gamma}^{\mu\nu}$, where
\begin{equation}
\delta\tl{\gamma}^{\mu\nu}=\ptl^{\mu}\tl{\xi}^{\nu}+
\ptl^{\nu}\tl{\xi}^{\mu}+c\eta^{\mu\nu}\,\ptl_{\alpha}\tl{\xi}^{\alpha},
\label{29}
\end{equation}
$c\ne-1/2$, and $\tl{\xi}^{\nu}$ is an arbitrary vector density of weight
$l$. It proves expedient to set $c=-l$. For any $S_{\rf}$ invariant in this
sense, a certain linear combination of the free-field equations is
identically divergenceless:
\begin{equation}
\ptl^{\mu}\left(\frac{\delta S_{\rf}}{\delta\tl{\gamma}^{\mu\nu}}-
\frac l2\eta_{\mu\nu}\eta^{\sigma\alpha}\frac{\delta S_{\rf}}
{\delta\tl{\gamma}^{\sigma\alpha}}\right)=0.
\label{30}
\end{equation}
This is the generalized Bianchi identity for the free-field theory. Local
energy--momentum conservation, which holds with the use of the
Euler--Lagrange equations for the gravity $\tl{\gamma}^{\mu\nu}$ and matter
$u$, can be written as
\begin{equation}
\ptl^{\nu}\left(\frac{\delta S}{\delta\tl{\eta}^{\mu\nu}}-
\frac l2\tl{\eta}^{\alpha\beta}\tl{\eta}_{\mu\nu}
\frac{\delta S}{\delta\tl{\eta}_{\alpha\beta}}\right)=0.
\label{31}
\end{equation}
We write the universal-coupling postulate in the form
\begin{equation}
\frac{\delta S}{\delta\tl{\gamma}^{\mu\nu}}=
\frac{\delta S_{\rf}}{\delta\tl{\gamma}^{\mu\nu}}+
\lambda\frac{\delta S}{\delta\tl{\eta}^{\mu\nu}}
\label{32}
\end{equation}
(the reason for choosing of the sign of the term containing the
energy--momentum tensor soon becomes clear). We obtain $S$, changing from
$\tl{\gamma}^{\mu\nu}$ and $\tl{\eta}^{\mu\nu}$ to the bimetric variables
$\tl{g}^{\mu\nu}$ and $\tl{\eta}^{\mu\nu}$, where
\begin{equation}
\tl{g}^{\mu\nu}=\tl{\eta}^{\mu\nu}+\lambda\tl{\gamma}^{\mu\nu}.
\label{33}
\end{equation}
The coefficient of $\lambda\tl{\gamma}^{\mu\nu}$ is chosen such that the
covariant and contravariant cases, as far as is easily achieved, define the
gravitational potential similarly. In particular, to the linear order in the
potential in Cartesian coordinates, the traceless part of the gravitational
potential $\gamma$ has the same observable significance, whether it is the
density difference between the covariant curved and flat and curved metrics
or the density difference between the contravariant flat and curved
metrics.\footnote{If $g_{\mu\nu}=\eta_{\mu\nu}-\lambda\gamma_{\mu\nu}$, then
the inverse metric yields an infinite series expansion (at least formally),
whose first term has a sign different from what a naive index raising might
suggest: $g_{\mu\nu}=\eta_{\mu\nu}+\lambda\gamma_{\mu\nu}+\dotsb$ implies
that $g^{\mu\nu}-\eta^{\mu\nu}\approx\lambda\gamma^{\mu\nu}$, not
$-\lambda\gamma_{\mu\nu}$. If the exact relation $g^{\mu\nu}-\eta^{\mu\nu}=
\lambda\psi^{\mu\nu} $ holds (the sign of the $\lambda$ term is important),
then $\psi_{\mu\nu}\approx\gamma_{\mu\nu}$. Thus, the meaning of the
gravitational potential is insensitive to a sign change, and it is therefore
easier to compare the various massive theories.} Equating coefficients of the
variations gives
\begin{equation}
\frac{\delta S}{\delta\tl{\eta}^{\mu\nu}}\bigg|\tl{\gamma}=
\frac{\delta S}{\delta\tl{\eta}^{\mu\nu}}\bigg|\tl{g}+
\frac{\delta S}{\delta\tl{g}^{\mu\nu}}
\label{34}
\end{equation}
and
\begin{equation}
\frac{\delta S}{\delta\tl{\gamma}^{\mu\nu}}=
\lambda\frac{\delta S}{\delta\tl{g}^{\mu\nu}},
\label{35}
\end{equation}
whence we obtain
\begin{equation}
\frac{\delta S_{\rf}}{\delta\tl{\gamma}^{\mu\nu}}=
-\lambda\frac{\delta S}{\delta\tl{\eta}^{\mu\nu}}\bigg|\tl{g}.
\label{36}
\end{equation}
The use of the generalized Bianchi identity implies that $S$ splits into a
component $S_1[\tl{g}^{\mu\nu},u]$ and a component $S_2$ that takes
form~\eqref{16}. The quantity $S_2$ contains all the ineliminable dependence
on the background metric and does not contribute to the field equations. The
simplest choice of $S_1$ gives the Hilbert action for Einstein's equations
and a cosmological constant, as in the covariant case. The specific choice
from the allowed values of $l$ makes no difference in the massless case.

\section{Derivation for a contravariant tensor density potential: Massive
case}
\label{sec5}

The choice of the density weight $l$ makes a difference in the massive
generalization of this contravariant derivation, much as in the covariant
case. The FMS--Logunov theory turns out to be the $l{=}1$ contravariant
universally coupled massive theory. While in some clear senses the $l{=}1$
theory is the best of the contravariant massive theories, it is not the only
such theory. The free-field action $S_{\rf}$ again has two parts: the part
$S_{\rf0}$ that is mostly kinetic and has a local gauge symmetry and an
algebraic mass term $S_{\fm}$. The action $S$ of the full theory again splits
into two parts, $S_0$ and $S_{\ms}$.

Requiring $S_{\rf0}$ to change only by a boundary term under the
infinitesimal variation
\begin{equation}
\delta\tl{\gamma}^{\mu\nu}=\ptl^{\mu}\tl{\xi}^{\nu}+\ptl^{\nu}\tl{\xi}^{\mu}-
l\eta^{\mu\nu}\,\ptl_{\alpha}\tl{\xi}^{\alpha},
\label{37}
\end{equation}
where $l\ne1/2$, implies the identity
\begin{equation}
\ptl^{\mu}\left(\frac{\delta S_{\rf0}}{\delta\tl{\gamma}^{\mu\nu}}-
\frac l2\eta_{\mu\nu}\eta^{\sigma\alpha}\frac{\delta S_{\rf0}}
{\delta\tl{\gamma}^{\sigma\alpha}}\right)=0.
\label{38}
\end{equation}
We again postulate the universal coupling in form~\eqref{32}. We change to
the bimetric variables $\tl{g}^{\mu\nu}$ and $\tl{\eta}^{\mu\nu}$. Letting
the new mass terms and the terms previously present agree separately, we find
that the mass terms satisfy
\begin{equation}
\frac{\delta S_{\fm}}{\delta\tl{\gamma}^{\mu\nu}}=
-\lambda\frac{\delta S_{\ms}}{\delta\tl{\eta}^{\mu\nu}}\bigg|\tl{g}.
\label{40}
\end{equation}
The action $S_{\fm}$ is chosen to be quadratic in the gravitational potential
and to satisfy
\begin{equation}
\frac{\delta S_{\fm}}{\delta\tl{\gamma}^{\mu\nu}}=
a\sqrt{-\eta}\,\tl{\gamma}^{\alpha\beta}
(\tl{\eta}_{\alpha\mu}\tl{\eta}_{\beta\nu}+
b\tl{\eta}_{\alpha\beta}\tl{\eta}_{\mu\nu}).
\label{40new}
\end{equation}
The quantity $S_{\ms}$ is naturally chosen in the form
\begin{equation}
S_{\ms}=
\int d^4x\,(p\tl{g}^{\alpha\beta}\tl{\eta}_{\alpha\beta}+q)\sqrt{-\eta}
\label{40a}
\end{equation}
for unspecified $p$ and $q$. Matching the coefficients of several terms gives
\begin{alignat*}{2}
&q=\frac{-(2-4l)a(1+4b)}{\lambda^2},&\qquad
&p=\frac{ab(2-4l)}{\lambda^2},\qquad p=\frac{a}{\lambda^2},
\\[2mm]
&b=-\frac1{4l-2},&\qquad&q=\frac{2 a (2l-3)}{\lambda^2}.
\end{alignat*}
Requiring the zeroth-order algebraic term in $S$ to vanish gives
$$
B=\frac{\Lambda}{16\pi G}+\frac{a(1-2l)}{\lambda^2}.
$$
Requiring the first-order algebraic term to vanish, after some algebra, gives
$\Lambda=-a(1-2l)/2$. The second-order term agrees with the free-field mass
term, as could be hoped. The spin-2 graviton mass $m$ is given by $a=-m^2$.

Combining all these results, we obtain the total massive action $S$, which
depends on the spin-2 graviton mass $m$ and the parameter $l$ controlling the
relative mass of the spin-0 graviton:
\begin{align}
S={}&\frac1{16\pi G}\int d^4x\,\sqrt{-g}\,R(g)+S_{\mtr}[\tl{g}^{\mu\nu},u]+{}
\nonumber
\\[2mm]
&+\frac12\int d^4x\,R_{\mu\nu\rho\sigma}(\eta)\cM^{\mu\nu\rho\sigma}
[\tl{\eta}^{\mu\nu},\tl{g}^{\mu\nu},u]+\int d^4x\,\alpha^{\mu},_{\mu}-{}
\nonumber
\\[2mm]
&-\frac{m^2}{16\pi G}\int d^4x\,\biggl[-\sqrt{-g}(2l-1)+\sqrt{-\eta}(2l-3)+
\frac12\sqrt{-\eta}\,\tl{g}^{\mu\nu}\tl{\eta}_{\mu\nu}\biggr],
\label{41}
\end{align}
where $l\ne 1/2$. The empirically doubtful Fierz--Pauli mass term is not
among the universally coupled theories considered in this paper. All these
theories are contained in the OP 2-parameter family.

While we could use some contravariant tensor (or tensor density) to find the
Euler--Lagrange equations, the equations are more easily compared with those
previously found if the metric $g_{\mu\nu}$ is used. The resulting equations
are
\begin{equation}
\frac{\delta S}{\delta g_{\mu\nu}}=-\frac{\sqrt{-g}}{16\pi G}G^{\mu\nu}-
\frac{m^2}{16\pi G}\biggl[\frac{1-2l}{2}\sqrt{-g}\,g^{\mu\nu}+
\frac{(\sqrt{-g}\,)^l\eta_{\alpha\beta}}{4(\sqrt{-\eta}\,)^{l-1}}
(lg^{\alpha\beta} g^{\mu\nu}-2g^{\mu\alpha}g^{\nu\beta})\biggr]+
\frac{\delta S_{\mtr}}{\delta g_{\mu\nu}}.
\label{42}
\end{equation}
Linearizing these theories shows that the spin-0 field is not a tachyon if
and only if $1/2<l\le3/2$. For $l=1$, the spin-0 ghost has the same mass as
the spin-2 degrees of freedom, and these field equations, which correspond to
the FMS--Logunov theory, are hence the cleanest in the set. At the linear
level, the $(0,2)$ theory of weight $-l$ is identical to the $(2,0)$ theory
of weight $l+1$. Hence, for the $(2,0)$ theory of weight $l+1$, the ratio of
the spin-0 mass $m_0$ to the spin-2 mass $m_2$ is given by
$$
\frac{m_0^2}{m_2^2}=\frac{-4(l-1)^2+1}{2(l-1)^2+1}.
$$
The mass is an even function of $l-1$. For $l=3/2$, the spin-0 degree of
freedom is massless, and $\sqrt{-\eta}$ is absent from the mass term. As
$l\to1/2$, the scalar again becomes light, and the coefficient of $\sqrt{-g}$
tends to $0$, although the value $l=1/2$ is forbidden. Between these massless
endpoints, the spin-0 degree of freedom becomes heavier. At the midpoint
$l=1$ (the FMS--Logunov theory), the spin-0 graviton has the same mass as the
spin-2 graviton, giving a simple form of the wave equation for the linearized
massive Einstein equations. Hence, the spin-0 ghost is never heavier than the
spin-2 degrees of freedom for this family of theories, although such can
occur for the larger family of OP massive theories~\cite{7}.\footnote{It is
reassuring that in the conformally flat special case $g_{\mu\nu}=
\phi\eta_{\mu\nu}$, the $(2,0)$ theory of weight $l+1$ and $(0,2)$ theory of
weight $-l$ coincide.}

\section{Massive gravities and experiment}
\label{sec6}

In both experimental~\cite{41} and theoretical contexts, it is common to
speak of {\sl the} mass of {\sl the} graviton, as if all gravitons must have
the same mass. While all gravitons do have the same mass in the most famous
spin-2--spin-0 massive gravity (developed by FMS and studied by Logunov and
collaborators), the existence of the OP theories shows that massive gravity
has two mass parameters that should be tested experimentally. It would be
worthwhile to ascertain to what degree the tacit assumption of equal spin-2
and spin-0 masses is actually used in finding experimental bounds on massive
gravity. Astrophysical tests for changes in the behavior of the degrees of
freedom present in massless general relativity primarily bound the spin-2
mass. The empirical bounds on the spin-2 graviton mass are so tight that the
spin-2 part of the mass term is empirically negligible except in strong
fields or over cosmic distances~\cite{20},~\cite{21}; these regimes are also
those investigated by Logunov and collaborators. The flexibility in the
spin-0 mass in the massive theories derived here opens some phenomenological
opportunities by increasing the range of the spin-0 repulsion that
counterbalances some of the spin-2 attraction. The larger family of OP
massive theories permits either longer or shorter range for the spin-0
repulsion compared with the spin-2 attraction. Babak and Grishchuk recently
investigated a similar phenomenological flexibility~\cite{21}. They noted
that their massive spin-2, massless spin-0 special case agreed with general
relativity in cosmological contexts because of the high degree of symmetry.
Hence, cosmological limits on the graviton mass(es) primarily bound the
spin-0 mass. While the OP theories have nonlinear mass terms motivated from
first principles in contrast to Babak and Grishchuk's linear mass terms
motivated by mathematical simplicity, similar qualitative behaviors of the
two kinds of two-parameter massive gravities can be expected outside highly
nonlinear regimes.

In view of the tight empirical bounds on the graviton masses, the observable
consequences of a mass term are rather difficult to detect. But there are two
important theoretical issues that arise at the classical level for massive
variants of Einstein's equations. The first is the well-known question of
stability, positive energy, etc., given the wrong-sign spin~0. The second is
the question of causality: massive gravity is a special relativistic field
theory with $\eta_{\mu\nu}$ observable, but there is reason to fear that
field propagation might violate causality by having the light cone of the
effective metric $g_{\mu\nu}$ leak outside that of $\eta_{\mu\nu}$. We now
turn to these issues.

\section{Positive energy}
\label{sec7}

It has long been argued that massive variants of Einstein's equations pose
the unpleasant dilemma~\cite{24}, \cite{42} that either the mass term is of
the Fierz--Pauli form with $5\infty^3$ degrees of freedom (pure spin~2) and
is empirically falsified by having a discontinuous massless limit or the
theory has $6\infty^3$ degrees of freedom including a wrong-sign spin~0 (a
spatial scalar density) and instability arises after linearization. The
former problem is the van Dam--Veltman--Zakharov discontinuity~\cite{43},
about which a large literature has appeared in the last few years after a
long period of relative quiet. More relevant for our purposes is whether the
massive theories with $6\infty^3$ degrees of freedom (spins 2 and 0) are
unstable.

Contrary to widely held views, Visser argued that the massive theories with
$6\infty^3$ degrees of freedom might well be stable~\cite{20}. More recently,
Babak and Grishchuk argued that such theories actually are stable~\cite{21}.
Concerning the specific case of the FMS theory, the authors of the theory
were themselves unconvinced of instability~\cite{23}, although they did not
follow up on the matter after arguments for instability were published. In
the middle to late 1980s, Logunov and collaborators (such as Loskutov and
Chugreev) adopted the FMS theory as the massive version of the relativistic
theory of gravity. They argued that this specific theory might well be
stable~\cite{11}, linearization arguments notwithstanding. More compellingly,
Loskutov calculated the gravitational radiation from a bounded source and
concluded that it is in fact positive definite~\cite{44}, even though the
theory has a wrong-sign spin-0 component. It is curious that this conclusion
has received so little response.

While the question of positive energy (or positive mass, as is often said in
a gravitational context) has not been settled with a favorable outcome (in
the sense of a general proof that all exact solutions satisfying certain
energy conditions have positive mass), neither do the arguments for
instability from linearization seem compelling. It is useful to show using a
Hamiltonian formalism that linearization is untrustworthy because it
essentially changes the form of the Hamiltonian constraint such that
instability becomes more plausible than is true for the exact nonlinear
theory. Although the wrong-sign spin~0 is present in the nonlinear theory,
its kinetic energy is related to that of the right-sign degrees of freedom
such that it can easily radiate only together with the positive energy
degrees of freedom. Precisely this feature is lost upon linearization. This
important feature of the Hamiltonian constraint depends essentially on a
cubic term in the Hamiltonian density and hence on a quadratic term in the
field equations.

In the case of the FMS--Logunov contravariant weight-1 theory, the field
equations impose a lower bound on the Hamiltonian density a little below
zero, in contrast to boundary terms (which should be annihilated, at least in
static contexts, because of the exponential Yukawa falloff of the
gravitational potential of bounded systems). Discarding unimportant terms and
factors from action~\eqref{41}, we obtain the Lagrangian density
\begin{equation}
\cL=\sqrt{-g}R(g)-m^2\biggl(-\sqrt{-g}-\sqrt{-\eta}+
\frac12\sqrt{-g}\,g^{\mu\nu}\eta_{\mu\nu}\biggr).
\label{43}
\end{equation}
We use the ADM (3$+$1)-dimensional split\footnote{The ADM split is a
noncovariant (3$+$1)-representation of the metric tensor of Riemannian space
first proposed by Arnowitt, Deser, and Misner.}~\cite{31} and choose
coordinates (Cartesian, spherical, or the like) such that $\eta_{00}=-1$ and
$\eta_{0i}=0$. The curved metric $g_{\mu\nu}$ is then expressed in terms of a
lapse function $N$ relating the effective proper time to the coordinate time,
a shift vector $\beta^i$ expressing how the spatial coordinate system appears
to shift among the various time slices, and a curved spatial metric $h_{ij}$
with the inverse $h^{ij}$ and determinant $h$. Letting $g^{\mu\nu}$ be the
inverse curved metric as usual, we have $g^{00}=-N^{-2}$ (the inverse metric
being most convenient here), $g_{ij}=h_{ij}$, and $g_{0i}=h_{ij}\beta^j$. The
indices for three-dimensional quantities are raised and lowered with
$h_{ij}$. Dropping a divergence from the Hilbert-like action above, we have
the FMS massive version of the standard (3$+$1)-dimensional Lagrangian
density
\begin{equation}
\cL=N\sqrt{h}\biggl[{}^3\!R+K_{ab}K^{ab}-K^2+
m^2\biggl(1-\frac{h^{ij}\eta_{ij}}{2}\biggr)\biggr]+
m^2\biggl[\sqrt{-\eta}+\frac{\sqrt{h}}{2N}(\eta_{ij}\beta^i\beta^j-1)\biggr].
\label{44}
\end{equation}
Hereafter, we drop the superscript on ${}^3\!R$.

Defining canonical momenta as usual, we obtain the usual results
\begin{equation}
\pi^{ij}=\frac{\ptl\cL}{\ptl h_{ij,0}}=\sqrt{h}(K^{ij}-h^{ij}K),\qquad
P_i=\frac{\ptl \cL}{\ptl\beta^i_{,0}}=0,\qquad
P=\frac{\ptl \cL}{\ptl N_{,0}}=0.
\label{45}
\end{equation}
The four vanishing canonical momenta are called primary constraints in the
context of constrained dynamics~\cite{45}.

Performing the generalized Legendre transformation and using the primary
constraints gives the canonical Hamiltonian density
\begin{equation}
\cH=N\biggl[\cH_0+m^2\sqrt{h}\biggl(\frac12 h^{ij}\eta_{ij}-1\biggr)\biggr]+
\beta^i \cH_i-m^2\sqrt{-\eta}+
\frac{m^2\sqrt{h}}{2N}(1-\eta_{ij}\beta^i\beta^j),
\label{46}
\end{equation}
where, as usual,
$$
\cH_0=\frac1{\sqrt{h}}\biggl(\pi^{ij}\pi_{ij}-
\frac12\pi^2\biggr)-\sqrt{h}\,R,\qquad\cH_i=-2D_j\pi^j_i,
$$
and $D_j$ is the three-dimensional torsion-free covariant derivative
compatible with $h_{ij}$. For $m=0$, we recover the usual form that is purely
a sum of constraints, but $m\ne0$ destroys that form and leads to six, not
two, degrees of freedom. We note that we have retained the zeroth-order term
$-m^2\sqrt{-\eta}$, and Minkowski space--time hence has zero energy, as it
should. Boulware and Deser omitted this term~\cite{24}. Varying the lapse $N$
and shift vector $\beta^i$, we obtain the secondary constraints, namely, the
modified Hamiltonian constraint
\begin{equation}
\frac{\ptl \cH}{\ptl N}=\cH_0+m^2\sqrt{h}
\biggl(-1+\frac12h^{ij}\eta_{ij}\biggr)-
\frac{m^2\sqrt{h}}{2N^2}(1-\eta_{ij}\beta^i\beta^j)=0
\label{47}
\end{equation}
and the modified momentum constraint
\begin{equation}
\frac{\ptl\cH}{\ptl\beta^i}=\cH_i-\frac{m^2\sqrt{h}}{N}\eta_{ij}\beta^j=0.
\label{48}
\end{equation}
These constraints are second-class~\cite{46}. As Boulware and Deser pointed
out, we can use these relations to eliminate the lapse and shift from the
Hamiltonian density to obtain a partly on-shell Hamiltonian density purely in
terms of the true degrees of freedom and their momenta:
$$
\cH=\sqrt{2m^2\sqrt{h}\biggl[\cH_0+m^2\sqrt{h}\biggl(\frac{h^{ij}\eta_{ij}}2-
1\biggr)\biggr]+\cH_i \cH_j\eta^{ij}}-m^2\sqrt{-\eta}.
$$
Expressing the lapse in terms of the true degrees of freedom, we obtain
\begin{equation}
N^2=\frac{h m^4}{2 m^2\sqrt{h}
\bigl[\cH_0+m^2\sqrt{h}(h^{ij}\eta_{ij}/2-1)\bigr]+\cH_i \cH_j\eta^{ij}}.
\label{49}
\end{equation}
Changing the variables from the lapse $N$ to the recently popular ``slicing
density" $\alpha=N/\sqrt{h}$~\cite{47} allows writing the on-shell
Hamiltonian density as $\cH=m^2(1/\alpha-\sqrt{-\eta}\,)$. Boundary terms
have been omitted, but they vanish in some cases because of the Yukawa
fall-off for localized sources.

We now consider the linearization of the exact Hamiltonian density. The
slicing density $\alpha=N/\sqrt{h}$ has the virtue of reducing the number of
radicals in the off-shell Hamiltonian density and also giving the on-shell
Hamiltonian density a simple form. Before linearization, we have
\begin{align}
\cH={}&\alpha\biggl[\pi^{ij}\pi^{kl}
\biggl(h_{ik}h_{jl}-\frac12h_{ij}h_{kl}\biggr)-
hR+m^2 h\biggl(-1+\frac12h^{ij}\eta_{ij}\biggr)\biggr]+{}
\nonumber
\\[2mm]
&+\beta^i \cH_i-m^2\sqrt{-\eta}+
\frac{m^2}{2\alpha}(1-\eta_{ij}\beta^i\beta^j).
\label{50}
\end{align}
We now let $\eta_{ij}=\delta_{ij}$, $h_{ij}=\delta_{ij}+\phi_{ij}$, and
$\alpha=1+a$. We are only interested in the kinetic term. The quintic and
quartic terms containing $a\Pi^2\phi^2$, $\Pi^2\phi^2$, and $a\Pi^2\phi$
(indices suppressed) are dispensable, but dropping the cubic term
$a(\pi^{ij}\pi^{ij}-\pi^{ii}\pi^{jj}/2)$ creates serious problems connected
with the term $-\pi^{ii}\pi^{jj}/2$ that did not arise in the exact theory.
The kinetic term containing the worrisome wrong-sign scalar $-\pi^2/2$ is
located in the Hamiltonian constraint of the exact theory, where its ability
to do damage is mitigated, but after linearization, the $-\pi^2/2$ term leads
to troubles, such as by radiating arbitrarily much negative energy away or by
permitting the radiation of arbitrarily much positive energy, leading to
instability. If we had the cubic terms such as $a(\pi^{ij}\pi^{ij}-
\pi^{ii}\pi^{jj}/2)$ and perhaps its spatial derivative analogue, then the
approximate Hamiltonian constraint would still substantially resemble the
exact form and might behave better.

Boulware and Deser~\cite{24}, noting the on-shell square root form of $\cH$
above (and lacking the zeroth-order term), commented that ``the Hamiltonian
form (in terms of 6 degrees of freedom)\,\dots\,appears to be positive
definite. Since in addition, the linearized approximation here corresponds to
a scalar-ghost admixture, and so gives the linearized Einstein interaction in
the weak-field limit, it would seem that this model has at least two
improvements over [the empirically doubtful Fierz--Pauli theory and
generalizations thereof]: Its energy is positive and it has correct
linearized behavior. However, it is unacceptable: The vacuum is not a local
minimum, but only a saddle point, as may be seen by considering equilibrium
(static) configurations, or simply expanding $H$ to quadratic order, where it
is found to agree with the linearized (ghost) version $H$. That for
appropriate excitations the quadratic part of $H$ can be negative may seem
irrelevant in view of the apparent positivity of the complete $H$\,$\dots$.
Unfortunately, the argument of the square root is not intrinsically
positive\,\dots\,even though its positivity is required for the theory to
make sense, i.e., for $N^2$ to be positive\,\dots\,(otherwise, the effective
Riemannian metric `seen' by matter will become pathological). Therefore one
would have to {\sl impose} that the excitations respect this requirement,
i.e., cut off arbitrarily those modes which take $H$ below its vanishing
vacuum ($g=\eta$) state value. Instability near vacuum ($g\simeq\eta$) is the
reason for rejecting this and other models whose quadratic mass is not of
Pauli--Fierz form."

This is a puzzling argument because $N\le0$ gives a singularity; there is
hence no need to ``impose" $N>0$ by hand. A remaining question is whether the
theory hits $N=0$ so often that singularities form in mundane contexts. No
argument to that effect has been given, while the fact that $N\to0$ implies
gravitational time dilation suggests that $N$ has little tendency to
vanish~\cite{48}. There seems to be no need for $\cH$ to be positive
everywhere as long as $H=\int\cH\,d^3x$ is positive (or nonnegative) for some
suitable boundary conditions. The restoration of the negative zeroth-order
term to $\cH$ implies that $\cH$ is bounded from below but not by zero.

There are currently (to our knowledge) no known solutions of the nonlinear
field equations, exact or numerical, of FMS--Logunov or any other ``ghost"
theory of the families considered here that have negative total energy. The
same is true for solutions that indicate instability by radiating negative
net energy. Given the need for a nonperturbative treatment, the question of
stability might best be resolved with the help of numerical relativists. It
suffices to work in spherical symmetry, where the wrong-sign field can
radiate but most of the right-sign fields cannot.

\section{Causality}
\label{sec8}

Given that massive gravities are considered in Minkowski space--time with an
observable background metric $\eta_{\mu\nu}$, a further issue worth
considering is whether the null cone of the flat background metric is a bound
of the effective curved metric $g_{\mu\nu}$. The flat metric is observable,
and violation of the null cone of $\eta_{\mu\nu}$ hence implies backwards
causation in some Lorentz frames, which is usually rejected. Causality for
higher-spin theories has already caused trouble in the case of spin-3/2
fields. Velo and Zwanziger~\cite{49} concluded that the ``main lesson to be
drawn from our analysis is that special relativity is not automatically
satisfied by writing equations that transform covariantly. In addition, the
solutions must not propagate faster than light."

The argument has been made that massive gravity leads to causality violation
in the sense of special relativity (this is relevant because of the
observable flat metric~\cite{13},~\cite{50}). As Chugreev rightly notes, the
static field of sources, if any, must be taken into account; for cosmological
models, the presence of matter everywhere might suffice to preserve
causality~\cite{51}. But surely it is a contingent rather than necessary
truth that the universe is filled with matter everywhere.\footnote{Indeed,
that claim is better regarded as a convention and not a fact, even in
cosmology intending to describe the actual world~\cite{6}.} Gravitational
radiation decays as $1/r$, and the static field due to localized sources
decays as $e^{-mr}$. Therefore, the radiation eventually wins and threatens
the proper relation between the two null cones. Because we wish to regard
many solutions without matter everywhere and with gravitational radiation as
physically meaningful (although apparently not corresponding to the actual
world), some additional strategy for ensuring the correct relation between
the null cones of the two metrics is needed. The only option that comes to
mind is to install artificial gauge freedom, perhaps using parameterization
along the lines of~\cite{50},~\cite{52}, and then to use the same strategy
that we used to impose $\eta$-causality in the massless case~\cite{13}. The
resulting gauged massive gravity has a gauge transformation groupoid, not a
group. It is noteworthy that parameterization yields results that for the
lowest-order terms resemble Stueckelberg's trick for introducing gauge
freedom into massive Proca electromagnetism. Stueckelberg's trick has
sometimes been used in the lowest order in gravity~\cite{53}, but it remained
unclear what the generalization to nonlinear field equations might be. It
seems plausible that other methods for installing artificial gauge freedom,
such as the BFT procedure~\cite{54} or gauge unfixing~\cite{55}, ought to
give similar results, although we have not investigated those questions
carefully.

\subsection*{Acknowledgments}
One of the authors (J.~B.~P.) thanks Katherine Brading for the assistance
with the history of Einstein's use of energy conservation and related
principles in his quest for gravitational field equations. The correspondence
with Yu.~V.~Chugreev, Stanley Deser, A.~A.~Logunov, Thanu Padmanabhan, and
Matt Visser, not all of whom agree with everything said here, is gratefully
acknowledged.

\end{document}